\documentclass[pra,twocolumn,nofootinbib,floatfix,10pt]{revtex4-2}

\usepackage[T1]{fontenc}
\usepackage{amsmath}
\usepackage{amssymb}
\usepackage{wasysym}
\usepackage{graphicx}
\usepackage{color,soul}
\usepackage{physics}
\usepackage{siunitx}
\usepackage{dsfont}
\usepackage{float}
\usepackage[english]{babel}
\usepackage{blindtext}
\usepackage[english,nomargin,inline,marginclue,draft]{fixme}
\pdfpageheight\paperheight
\pdfpagewidth\paperwidth

\usepackage[colorlinks,linkcolor=blue,anchorcolor=blue,citecolor=blue,urlcolor=blue]{hyperref}

\fxusetheme{colorsig}
\FXRegisterAuthor{cg}{acg}{CG}  
\FXRegisterAuthor{th}{ath}{\color{blue}TH}  
\FXRegisterAuthor{ib}{aib}{\color{red}IB} 
\FXRegisterAuthor{sh}{ash}{\color{cyan}SH} 
\FXRegisterAuthor{db}{adb}{\color{green}DB} 
\FXRegisterAuthor{ps}{aps}{PS}
\makeatletter
\renewcommand*\FXLayoutInline[3]{%
  {\@fxuseface{inline}\ignorespaces{\color{fx#1}[#3: #2]}}}
\makeatother

\long\def\symbolfootnote[#1]#2{\begingroup%
\def\thefootnote{\fnsymbol{footnote}}\footnotetext[#1]{#2}\endgroup}

\def\nobreakbefore{%
  \relax\ifvmode\else
    \ifhmode
      \ifdim\lastskip > 0pt\relax
        \unskip\nobreakspace
      \else 
        \nobreakspace
      \fi
    \fi
  \fi
}
\let\oldcite\cite
\renewcommand\cite{\nobreakbefore\oldcite}

\begin{document}

\title{Rydberg atomic field-enhanced feed beyond electrically small antennas}

\author{Jia-Dou Nan$^{1,2,6,\textcolor{blue}{\star}}$}
\author{Qian-Qian Zhang$^{1,2,3,\textcolor{blue}{\star}}$}
\author{Hao-Yu Fan$^{3,\textcolor{blue}{\star}}$}
\author{Jin-Hang Fang$^{4}$}
\author{Luan-Yi Wei$^{4}$}
\author{Bang Liu$^{1,2,4}$}
\author{Kai Wang$^{3,\textcolor{blue}{\S}}$}
\author{Li-Hua Zhang$^{1,2,4,\textcolor{blue}{\ddagger}}$}
\author{Qing-Li Ma$^{3}$}
\author{Bo Du$^{5,\textcolor{blue}{\star\star}}$}
\author{Dong-Sheng Ding$^{1,2,4,6,\textcolor{blue}{\dagger}}$}
\author{Bao-Sen Shi$^{1,2,6}$}

\affiliation{$^1$Key Laboratory of Quantum Information, University of Science and Technology of China, Hefei, Anhui 230026, China.}
\affiliation{$^2$Anhui Province Key Laboratory of Quantum Network, University of Science and Technology of China, Hefei, Anhui 230026, China.}
\affiliation{$^3$College of Electronic Engineering, National University of Defense Technology, Hefei, Anhui 230037, China}
\affiliation{$^4$Davinci Quantum Technology Co., Ltd, Hefei, Anhui 230026, China}
\affiliation{$^5$Qianyuan Laboratory, Hangzhou, Zhejiang 310024, China.}
\affiliation{$^6$Hefei National Laboratory, University of Science and Technology of China, Hefei, 230088, China.}
\date{\today}

\symbolfootnote[1]{J.D.N., Q.Q.Z. and H.Y.F. contribute equally to this work.}
\symbolfootnote[4]{wkhy@ustc.edu.cn}
\symbolfootnote[3]{zlhphys@ustc.edu.cn}
\symbolfootnote[6]{dubo@zju.edu.cn}
\symbolfootnote[2]{dds@ustc.edu.cn}
\date{\today}

\maketitle
\textbf{The reception of electromagnetic signals becomes increasingly challenging at longer wavelengths, as conventional antenna feeds must scale with the operating wavelength, making it difficult to reconcile compactness with high sensitivity. Rydberg atoms enable direct field-to-optical conversion within deeply subwavelength sensing volumes, yet an effective receiving feed that couples incident fields into the atomic interaction volume remains underdeveloped. Here, we introduce a Rydberg atomic field-enhanced feed (RAFEF) that integrates electromagnetic field capture and enhancement with Rydberg-atomic transduction. We establish a physical model to quantify its receiving capability, \(\Phi\), that jointly accounts for sensitivity enhancement, receiving bandwidth and normalized three-dimensional volume, and experimentally obtain \(\Phi_{\mathrm{dB}}=70.81~\mathrm{dB}\). Compared with a conventional loop-antenna feed, the RAFEF provides a higher received SNR and overall receiving capability, while its tunable narrowband response enables frequency selectivity directly at the receiving-feed stage. We further demonstrate selective recovery of a target channel from simultaneously incident signals and reception of a \(93.5~\mathrm{MHz}\) FM broadcast after approximately \(10~\mathrm{km}\) of over-the-air propagation. These results establish the RAFEF as an atomic receiving-feed architecture that exploits localized electric-field transduction to extend Rydberg-atomic reception beyond conventional electrically small antenna feeds.}

\section*{Introduction}

Electromagnetic signal reception is essential for applications including remote sensing, spectrum monitoring and wireless communication~\cite{Hengyan2024,Andr2025,Yongjun2023,Liu2022rfrecognition,Meyer2023Demodulation}. At frequencies of hundreds of megahertz, however, conventional receiving antenna feeds must scale with meter-scale wavelengths, making it increasingly difficult to reconcile compact dimensions with high receiving sensitivity and usable bandwidth. Electrically small antennas based on passive resonances, tunable elements~\cite{Jiahao2023}, active matching networks and time-varying modulation~\cite{Arora2026}  have substantially extended the performance of compact receivers~\cite{Yu2022,Strachen2024}. Nevertheless, as the electrical size of a metallic antenna decreases, its interaction with the incident electromagnetic field and the extraction of the received signal through an electrical port become increasingly constrained by size--efficiency--bandwidth trade-offs and impedance-matching requirements~\cite{Chu1948,Sievenpiper2011,Gustafsson2018}. These constraints motivate receiving architectures that do not rely exclusively on the conventional paradigm of extracting RF power through an electrically small metallic antenna.

Rydberg atoms provide a distinct physical basis for such a receiving feed architecture by enabling electric-field-to-optical transduction within a deeply subwavelength sensing volume~\cite{Zhang2024Ultra-wide,Liu2023Electric,Zhang2022Comb,ArtusioGlimpse2022,Yuan2023,Fancher2021Sensors,Anderson2021}. Their large polarizabilities and optically accessible energy-level structures allow incident radio-frequency (RF) fields to shift, dress or mix Rydberg states, with the resulting response read out through electromagnetically induced transparency (EIT)~\cite{Han2023EIT}, Autler--Townes splitting~\cite{Li2023AT}, AC Stark shifts~\cite{Yang2024AC} or optical heterodyne detection~\cite{Manchaiah2026,Jing2020superheterodyne}. These capabilities have also enabled the measurement of practical RF signals~\cite{Prajapati2022,Deb2018Radio,Borwka2023wideband}, including those generated by electrical arcs~\cite{Wang2025} and partial discharges~\cite{Wang2025,
Wang2025CPB}. In contrast to a conventional antenna, for which the received electromagnetic field is converted into an electrical-port signal, a Rydberg-atomic sensor directly responds to the local RF electric field within the optically interrogated atomic volume~\cite{Schlossberger2024}. The sensing volume can therefore remain deeply subwavelength without requiring the atomic transduction element itself to scale with the RF wavelength~\cite{Rotunno2023}. This field-sensitive and spatially localized transduction provides a distinct route towards long-wavelength electromagnetic signal reception beyond the conventional electrically small antenna feeds.

Substantial efforts have consequently been devoted to improving the low-frequency response of Rydberg-atomic sensors. Atomic-control and optical-readout techniques~\cite{ding2022enhanced,Wang2026,Liu2022MHz}, including heterodyne readout and phase demodulation~\cite{Jin2025}, modulated auxiliary-field sensing, Floquet engineering~\cite{Zhao2023,Wilkason2022,Yin2022,zhang2024tunable}, Rabi matching~\cite{Yin2022} and high-angular-momentum-state preparation~\cite{Machu2026}, enhance the conversion of a local RF field into a measurable atomic or optical response~\cite{Prajapati2024VHF,Song2024Floquet}. These approaches primarily improve the atomic transduction after the electromagnetic field has reached the sensing volume. Resonant and guided electromagnetic structures, including split-ring and multilayer resonators~\cite{Holloway2022,Wan2026Resonator,Wan2025resonator}, cavities~\cite{Wu2022Design,Ma2022Measurement,Liu2025Cavity}, planar resonators~\cite{Richardson2026Extraction,Yang2024PETR}, waveguides~\cite{Meyer2021Waveguide} and photonic-crystal structures~\cite{Amarloo2025} have further been employed to enhance and confine RF fields around Rydberg-atomic vapor cells~\cite{Sandidge2024,Li2025CTER,Wu2026Tshape}. Recent implementations have also  incorporated electrical or structural tuning~\cite{Hu2022tunable} and integrated microwave structures~\cite{Yang2023resonator,Wu2024port} to improve operational flexibility and field enhancement~\cite{Yang2025BFRR,Fan2015Geometry}. These studies demonstrate that resonant electromagnetic structures can improve the local atom--field interaction. 

However, enhanced atomic response or local-field sensitivity alone does not constitute an effective receiving feed. The central receiving challenge is to capture the incident electromagnetic field and efficiently couple it into the finite atomic interaction volume, where the local RF electric field is transduced into an optical signal. This requirement differs from that of a conventional electrically small antenna, whose receiving response is extracted through an electrical port, whereas a Rydberg-atomic receiver responds to the local electric field experienced by the atoms. An effective atomic receiving feed must therefore integrate incident-field coupling and local-field enhancement with spatial overlap between the enhanced field and the atomic interaction volume. Such an integrated receiving-feed architecture, together with a framework for benchmarking its receiving performance against a conventional antenna feed, remains to be established. 

Here, we introduce a Rydberg atomic field-enhanced feed (RAFEF), implemented using a logarithmic-spiral (LS) resonant structure, that integrates incident-field coupling and concentration with localized Rydberg-atomic field-to-optical transduction. We establish a physical model for the RAFEF and define a receiving-capability metric \(\Phi_{\mathrm{RAFEF}}\), that jointly accounts for sensitivity enhancement, receiving bandwidth and normalized three-dimensional volume. Compared with a conventional loop-antenna feed, the RAFEF achieves a higher received SNR and overall receiving capability, while its intrinsically narrowband and tunable response provides frequency selectivity directly at the receiving-feed stage. We further demonstrate selective recovery of a target channel from simultaneously incident signals and reception of a \(93.5~\mathrm{MHz}\) FM broadcast after approximately \(10~\mathrm{km}\) of over-the-air propagation. These results establish the RAFEF as an integrated atomic receiving-feed architecture that exploits localized electric-field transduction to extend Rydberg-atomic reception beyond conventional electrically small antenna feeds.

    \begin{figure*}
        \centering
        \includegraphics[width=1\linewidth]{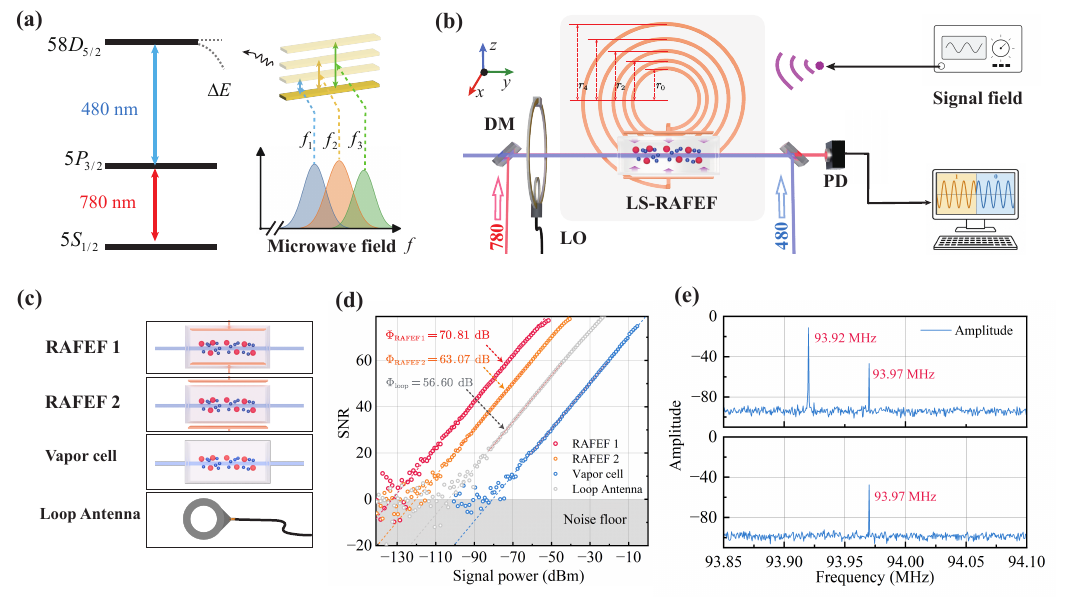}
        \caption{\textbf{RAFEF receiving architecture and signal-reception benchmark.} (a) Ladder-type energy-level scheme used for Rydberg excitation in Rb atoms. The 780-nm probe laser drives the \(5S_{1/2}\rightarrow5P_{3/2}\) transition, whereas the 480-nm coupling laser drives the \(5P_{3/2}\rightarrow58D_{5/2}\) transition. (b) Optical and radio-frequency (RF) configuration of the Rydberg atomic receiver. The counter-propagating 780-nm probe and 480-nm coupling beams pass through the atomic interaction region of the LS-RAFEF. (c) Schematics of the four receiver configurations. From top to bottom, RAFEF1 has its parallel-plate section located inside the vapor cell, whereas RAFEF2 has its parallel-plate section located outside the vapor cell. The other two configurations employ an unassisted vapor-cell receiver and a standalone loop-antenna receiver, respectively. (d) Sensitivity characterization of the four receiver configurations. The horizontal axis represents the applied signal power, and the vertical axis represents the measured SNR. The inset lists the receiving-capability metric \(\Phi\) evaluated for the corresponding receiving feeds. (e) Spectra of the RF fields used for the atomic and loop-antenna measurements. For the atomic receiver, the signal and local-oscillator fields are applied at 93.97 and 93.92 MHz, respectively, producing a 50-kHz beat signal. For the loop-antenna receiver, only the 93.97-MHz SIG field is directly applied, without an LO field.}
        \label{fig1}
    \end{figure*}

\section*{Physical Model}

\noindent\textbf{Physical framework of the RAFEF.} 
The RAFEF is designed to couple an incident radio-frequency (RF) field into the finite atomic interaction volume, where the resulting local electric field is transduced into an optical response. As illustrated in Fig.~\ref{fig1}(b), the implementation used here employs a logarithmic-spiral (LS) resonant structure to capture the incident RF field and concentrate the resulting electromagnetic response within the optically interrogated Rydberg-atomic volume. The spiral radius increases with the azimuthal angle according to a prescribed geometric scaling, such that the successive radii \(r_0,r_1,r_2,\ldots\) define a compact resonant path. The geometrical parameters are chosen to place the resonance within the target VHF range and to extend the enhanced electric field over the finite atomic interaction volume. In this way, the LS structure provides the electromagnetic implementation of the RAFEF, whereas the receiving response is determined by the coupling between the incident field, the resonant field distribution and the spatially localized atomic transduction.

To quantify the electromagnetic coupling provided by the RAFEF, we first consider the RF field experienced by the atomic ensemble. Rather than using the maximum field enhancement at a single spatial location, we define a volume-averaged field-enhancement factor,
\begin{equation}
\eta_E=\frac{\frac{1}{V_a}\int_{V_a}\left|E_{\mathrm{RAFEF}}\mathbf{(r)}\right|^2dV}{\left|E_{inc}\right|^2},
\end{equation}
where \(E_{\mathrm{RAFEF}}(\mathbf{r})\) is the spatial distribution of the RF electric field in the presence of the RAFEF, \(E_{\mathrm{inc}}\) is the incident electric field without field enhancement, and \(V_a\) denotes the effective atomic interaction volume defined by the spatial overlap between the RF field and the optical interrogation region. The quantity \(\eta_E\) therefore characterizes the electromagnetic coupling of the incident field to the finite atomic volume, rather than the maximum local enhancement of the resonant structure.

The electromagnetic field concentration described by \(\eta_E\) is transduced through the Rydberg atoms and optical readout into an experimentally measurable receiving response. We quantify this receiver-level improvement using the sensitivity-enhancement factor 
\begin{equation}
G_{\mathrm{c}}=\frac{P_{\mathrm{min}}^{0}}{P_{\mathrm{min}}^{\mathrm{RAFEF}}},
\end{equation}
where \(P_{\mathrm{min}}^{0}\) and \(P_{\mathrm{min}}^{\mathrm{RAFEF}}\) are the minimum detectable RF powers of the atomic receiver without and with the RAFEF, respectively. Unlike \(\eta_E\), which describes electromagnetic field coupling within the atomic volume, \(G_{\mathrm{c}}\) is obtained from the measured receiving response and therefore incorporates the combined effects of field concentration, atomic transduction and optical readout. The two quantities thus describe complementary aspects of the RAFEF: \(\eta_E\) identifies the electromagnetic origin of the enhancement, whereas \(G_{\mathrm{c}}\) quantifies the experimentally observed improvement of the complete receiving chain. 
For comparison between receiving feeds employing different transduction mechanisms, we further introduce a generic receiving enhancement \(G_{\mathrm{rec}}\), defined relative to the corresponding reference receiving configuration under the measurement convention specified below. For the RAFEF, \(G_{\mathrm{rec}}=G_c\).

\begin{figure*}
        \centering
        \includegraphics[width=1\linewidth]{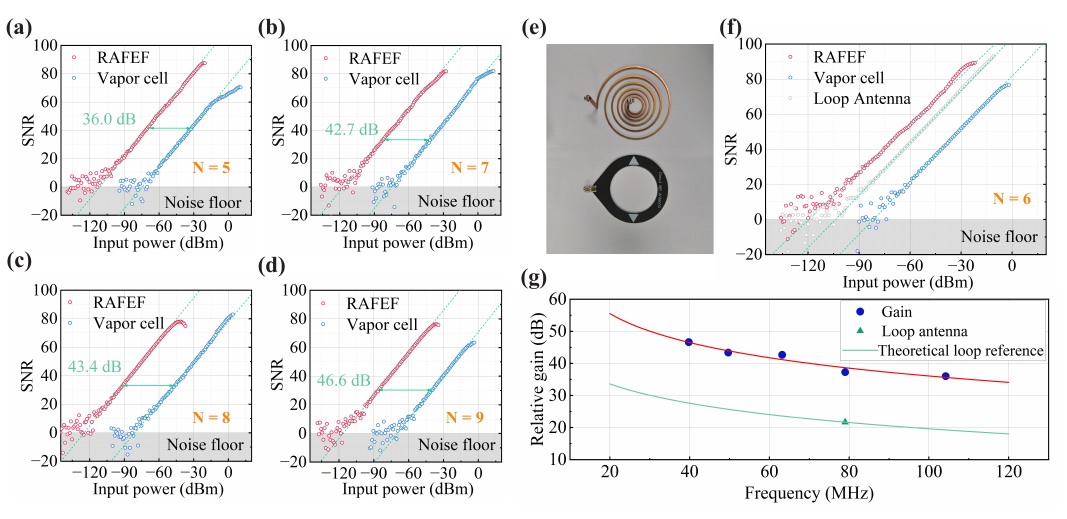}
        \caption{\textbf{Sensitivity enhancement of the RAFEF and comparison with a size-matched loop antenna.} (a)--(d) Measured SNR as a function of calibrated input power for the bare vapor cell and the RAFEFs with $N=5$, 7, 8 and 9, respectively. The linear fits are obtained from the signal-dominated regions, and the horizontal separations indicate the corresponding sensitivity enhancements. The shaded regions denote the noise floor. (e) Photograph of the $N=6$ RAFEF and the size-matched loop antenna. (f) Sensitivity comparison among the $N=6$ RAFEF, bare vapor cell and loop antenna. (g) Measured relative gain of the RAFEF as a function of resonant frequency. The red curve shows the fitted trend of the RAFEF results. The green triangle denotes the measured loop-antenna result at $78.99~\mathrm{MHz}$, and the green curve represents the theoretical loop reference calculated from geometrical area scaling and calibrated using this measurement.}
        \label{fig2-}
    \end{figure*}

Sensitivity enhancement alone, however, is insufficient to evaluate a compact resonant receiving feed, because stronger resonant enhancement is generally accompanied by changes in receiving bandwidth and electrical size. To compare receiving feeds with different transduction mechanisms and physical dimensions, we define a system-level receiving-capability metric as
\begin{equation}
\Phi_{\mathrm{RAFEF}}=\frac{G_{\mathrm{c}}\mathrm{FBW}}{V_{\mathrm{RAFEF}}/\lambda_0^3},
\end{equation}
where \(\mathrm{FBW}=\Delta f/f_0\) is the fractional 3-dB receiving bandwidth, \(f_0\) is the operating centre frequency, and \(\Delta f\) is the carrier-frequency bandwidth measured at a fixed tuning state, defined as the frequency interval over which the received response remains within 3 dB of its maximum value. This instantaneous bandwidth is distinct from the accessible tuning range obtained by reconfiguring the RAFEF resonance and from the intermediate-frequency response of the atomic receiver. The quantity \(V\) denotes the physical volume of the complete receiving feed defined using a common minimum-enclosing-volume convention. 
The corresponding values for the RAFEF and loop antenna are denoted by \(\Phi_{\mathrm{RAFEF}}\) and \(\Phi_{\mathrm{loop}}\), respectively.

For an atomic receiving feed, the relevant physical volume must include the complete electromagnetic--atomic interaction architecture rather than only the resonant metallic structure. We therefore define \(V_{\mathrm{RAFEF}}\) using the minimum enclosing sphere containing the vapor cell, field-enhancement structure and associated components,
\begin{equation}
V_{\mathrm{RAFEF}}=\frac{4}{3}\pi a_{\mathrm{RAFEF}}^3. 
\end{equation}
where \(a_{\mathrm{RAFEF}}\) is the equivalent radius of the complete
receiving architecture. The corresponding electrical size is given by \(ka_{\mathrm{RAFEF}}= {2\pi a_{\mathrm{RAFEF}}}/{\lambda_0}\), where \(\lambda_0\) is the free-space wavelength at the operating frequency. The normalized volume term can therefore be expressed as \({V_{\mathrm{RAFEF}}/\lambda_0^3} ={\left(k a_{\mathrm{RAFEF}}\right)^3}/6\pi^2\) which connects the physical volume of the atomic receiving feed to the electrical-size scaling commonly used for compact electromagnetic receiving structures.

The quantities \(\eta_E\), \(G_{\mathrm{c}}\) and \(\Phi_{\mathrm{RAFEF}}\) therefore characterize the RAFEF from electromagnetic coupling within the atomic volume to experimentally observed receiving enhancement and, finally, to the overall performance of the complete receiving feed. In particular, \(\Phi_{\mathrm{RAFEF}}\) enables the RAFEF to be evaluated as an integrated electromagnetic--atomic receiving feed rather than solely as a local field-enhancement structure, and provides the basis for the direct comparison with a conventional electrically small loop-antenna feed presented below.

\section*{Results}

\vspace{0.5\baselineskip}
\noindent\textbf{Receiving-performance advantage of the RAFEF.}
We first characterized the receiving performance of the RAFEF and benchmarked it against a conventional loop-antenna receiver. Figure~\ref{fig1} shows the two RAFEF configurations investigated in this work together with the corresponding receiving measurements. In RAFEF~1, the field-enhancement plates are positioned inside the vapor cell, whereas in RAFEF~2 they are positioned outside the cell. In both configurations, the logarithmic-spiral structure couples the incident RF field to the field-enhancement region, where the enhanced local electric field interacts with the Rydberg atomic ensemble and is transduced into an optical response. Further details of the structural geometry, simulated field distribution and logarithmic-spiral design rationale are provided in Supplementary Notes~1 and 2, Supplementary Figs.~1 and 2, and Supplementary Table~1.

    \begin{figure*}
        \centering
        \includegraphics[width=1\linewidth]{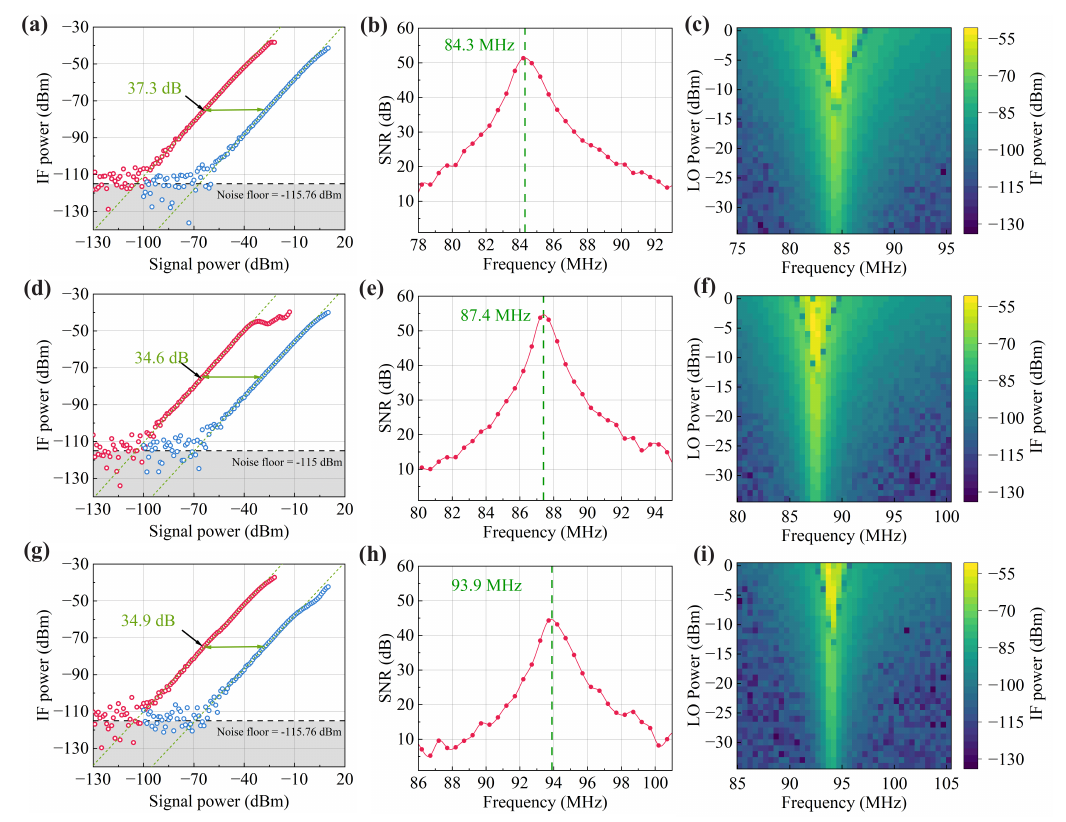}
       \caption{\textbf{Tunable receiving response of the RAFEF.} (a)--(c) Measurements with the resonant enhancement structure tuned to 84.3~MHz. (a) Received intermediate-frequency (IF) signal intensity as a function of signal-field power. Red and blue circles denote measurements with and without the resonant enhancement structure, respectively. The grey shaded region indicates the noise floor of approximately -115.76~dBm. (b) Signal-to-noise ratio (SNR) as a function of signal-field frequency. The red trace shows the measured SNR response, and the green vertical dashed line marks the resonance frequency of the resonant enhancement structure. (c) Heat map of the received IF signal intensity obtained by scanning the signal-field frequency and the local-oscillator (LO) field power. The result shows that an appropriate LO-field power is required to optimize the IF response, whereas excessive LO-field power drives the system into a nonlinear regime. (d)--(f) Corresponding measurements with the resonant enhancement structure tuned to 87.4~MHz. (g)--(i) Corresponding measurements with the resonant enhancement structure tuned to 93.9~MHz. The gain is extracted from the difference in signal-field power required for the received IF signal to reach the same noise-floor-limited detection threshold.}
        \label{fig3-}
    \end{figure*}

Figure~\ref{fig1}(d) compares the measured signal-to-noise ratio (SNR) as a function of signal-source power for RAFEF~1, RAFEF~2, the bare vapor-cell receiver and the conventional loop-antenna receiver. For a given SNR, RAFEF~1 requires the lowest signal-source power, followed by RAFEF~2 and the loop antenna, whereas the bare vapor-cell receiver requires the highest source power. The displacement of the RAFEF responses relative to the bare vapor-cell measurement quantifies the improvement introduced by the electromagnetic receiving feed. RAFEF~1 exhibits the larger receiving enhancement, consistent with the stronger overlap between the enhanced RF field and the optically interrogated atomic region indicated by the electromagnetic characterization.

The loop-antenna measurement provides a system-level experimental benchmark against a conventional receiving feed. Without external low-noise amplification in either receiver, RAFEF~2 reaches a given SNR at a lower signal-source power than the loop-antenna receiver (Fig.~\ref{fig1}(d)). Because the two receivers employ different transduction and detection mechanisms, this comparison is performed at the system level using their respective receiving and detection configurations. The corresponding loop-antenna response with additional low-noise amplification is provided for comparison in Supplementary Note~6 and Supplementary Fig.~4. To further account for receiving enhancement, instantaneous receiving bandwidth and physical volume, we evaluated the receiving-capability metric \(\Phi\) defined above. At the corresponding operating frequency, the values are \(\Phi_{\mathrm{RAFEF1}}=\mathrm{70.81~dB}\),
\(\Phi_{\mathrm{RAFEF2}}=\mathrm{63.07~dB}\), and \(\Phi_{\mathrm{loop}}=\mathrm{56.60~dB}\), as indicated in Fig.~\ref{fig1}(d). The higher \(\Phi\) values obtained for the RAFEF show that the receiving-performance advantage is retained when receiving enhancement, instantaneous bandwidth and normalized three-dimensional volume are considered jointly. The parameters and calculation procedure used to obtain these values are provided in Supplementary Note~7 and Supplementary Table~2. The measured 3-dB receiving bandwidth of the loop antenna is shown in Supplementary Fig.~3.

Having established the receiving-performance advantage at a representative operating frequency, we next examined RAFEF structures designed for different operating frequencies. Among the RAFEF structures investigated, five representative designs were selected for the measurements shown in Fig.~\ref{fig2-}. These structures have different logarithmic-spiral geometries, with turn numbers from \(N=5\) to \(N=9\), and span measured resonance frequencies from \(104.2\) to \(39.78~\mathrm{MHz}\). The corresponding device photographs and geometrical parameters are provided in Supplementary Fig.~2 and Supplementary Table~1. The source-power-dependent SNR measurements in Figs.~\ref{fig2-}(a--d) and 2(f) show substantial sensitivity enhancement relative to the bare vapor-cell receiver for each representative RAFEF structure.
The enhancement values extracted from these measurements are summarized in Fig.~\ref{fig2-}(g). Over the investigated frequency range, the five measured RAFEF values exhibit an approximately logarithmic frequency dependence, as indicated by the empirical fit. These measurements show that substantial receiving enhancement can be realized using RAFEF structures designed for different operating frequencies across the investigated portion of the VHF band.

For comparison, we calculated the frequency-dependent receiving enhancement of the conventional loop antenna using the theoretical model described in Supplementary Note~7. The calculated response is shown by the green curve in Fig.~\ref{fig2-}(g), together with an experimental loop-antenna measurement at \(78.99~\mathrm{MHz}\). The measured enhancement values of the five representative RAFEFs are higher than the calculated loop-antenna reference at their corresponding operating frequencies. Together with the direct experimental benchmark in Fig.~\ref{fig1}(d), these results extend the receiving-performance comparison from a representative operating
condition to RAFEF structures designed for different VHF operating frequencies.

\begin{figure*}
        \centering
        \includegraphics[width=1\linewidth]{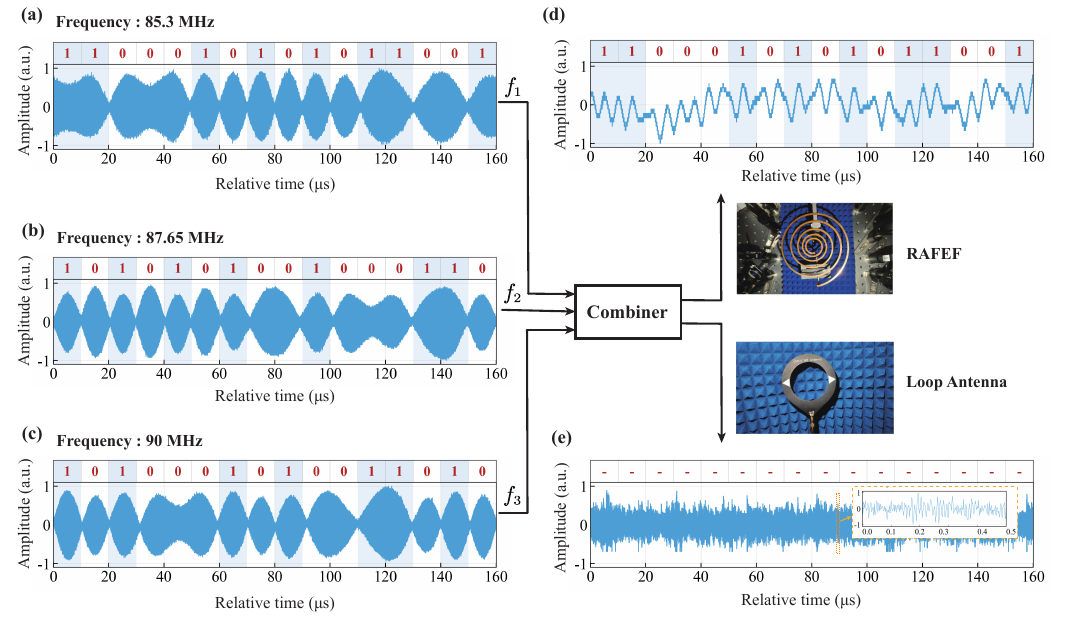}
        \caption{\textbf{Frequency-selectivity advantages of the RAFEF over a conventional loop antenna.} (a)-(c) Time-domain demonstration of frequency-selective reception. Three independently modulated signals at $f_1=85.3~\mathrm{MHz}$, $f_2=87.65~\mathrm{MHz}$, and $f_3=90~\mathrm{MHz}$ are combined and transmitted simultaneously to the RAFEF and the loop antenna. (d) When tuned to $f_1$, the RAFEF selectively recovers the data carried by the 85.3-MHz channel without additional post-processing. (e) In contrast, the broadband loop antenna receives a superposition of the three signals and does not isolate the target channel at the receiving-feed stage.}
        \label{fig4-}
    \end{figure*}


\vspace{0.5\baselineskip}
\noindent\textbf{Tunable narrowband reception and frequency selectivity.}
We next characterized the frequency-tunable receiving response of the RAFEF and examined how its narrowband resonance enables frequency-selective signal reception. The resonance frequency was adjusted by varying the plate spacing, while the atomic excitation and optical readout configurations were maintained unchanged. As shown in Fig.~\ref{fig3-}, three representative resonance frequencies of \(84.3\), \(87.4\), and \(93.9~\mathrm{MHz}\) were selected from the lower, intermediate, and upper portions of the accessible tuning range, respectively. The resonance frequency spans \(9.6~\mathrm{MHz}\), corresponding to a relative tunable range of \(10.35\%\).

Figures~\ref{fig3-}(a), (d) and (g) show the received intermediate-frequency (IF) power as a function of signal-source power at the three resonance frequencies. The red and blue data correspond to measurements with and without the RAFEF enhancement, respectively. Under the same acquisition conditions, the measured noise floor is approximately \(-115.76~\mathrm{dBm}\).  At each operating frequency, the RAFEF shifts the noise-floor-limited detection threshold towards lower signal-source powers, corresponding to threshold improvements of \(37.3\), \(34.6\), and \(34.9~\mathrm{dB}\) at \(84.3\), \(87.4\), and \(93.9~\mathrm{MHz}\), respectively. The threshold improvement therefore remains above \(34~\mathrm{dB}\) across the three representative tuning states. 

Figures~\ref{fig3-}(b), (e) and (h) show the frequency-dependent SNR for the three RAFEF configurations. For each state, the SNR reaches a maximum near the corresponding RAFEF resonance and decreases as the signal frequency is detuned from resonance. The peak SNR values are approximately \(50\), \(52\), and \(44~\mathrm{dB}\) at resonance frequencies of \(84.3\), \(87.4\), and \(93.9~\mathrm{MHz}\), respectively. As the RAFEF resonance is shifted across the tuning range, the frequency of maximum receiving response shifts accordingly, demonstrating that the narrowband receiving response can be repositioned in frequency.

Figures~\ref{fig3-}(c), (f) and (i) further show the dependence of the received IF power on signal frequency and local-oscillator (LO) power. At all three tuning states, the IF response initially increases with LO power and reaches its strongest response near \(-5~\mathrm{dBm}\), before deviating from the lower-power trend at higher LO powers. The similar behaviour observed across the three resonance frequencies identifies a consistent LO operating region for heterodyne reception over the measured tuning range. These measurements establish a tunable narrowband receiving response over \(84.3{-}93.9~\mathrm{MHz}\), corresponding to a \(9.6~\mathrm{MHz}\) tuning span and a relative tuning range of \(10.35\%\). The source-power threshold improvement remains above \(34~\mathrm{dB}\) across the three representative tuning states, while the frequency of maximum receiving response follows the tuned RAFEF resonance. The RAFEF therefore combines substantial receiving enhancement with a narrowband response that can be repositioned across the accessible frequency range.

    \begin{figure*}
        \centering
        \includegraphics[width=1\linewidth]{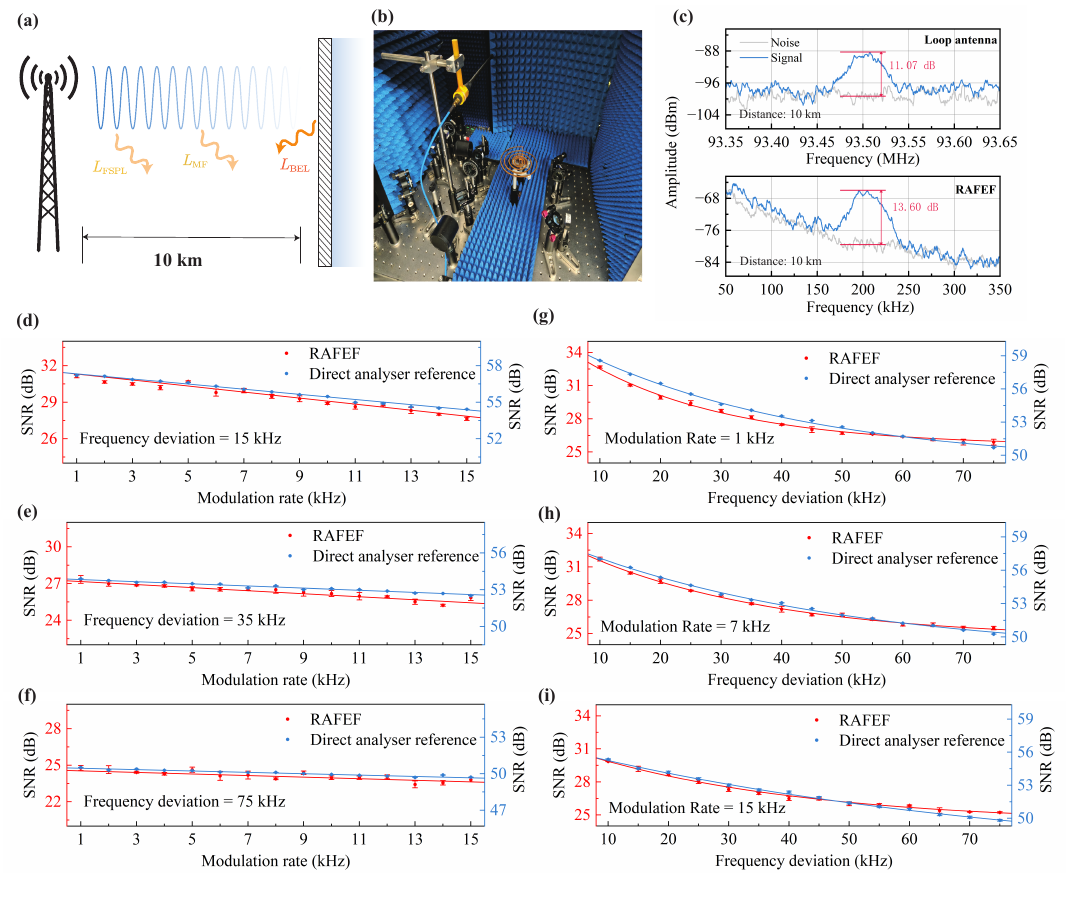}
       \caption{\textbf{FM signal reception and over-the-air validation of the RAFEF.} (a) Schematic of FM broadcast reception. The broadcast transmitter is located approximately 10~km from the measurement site, and the receiver is placed indoors. During propagation, the broadcast signal experiences free-space path loss ($L_{\rm FSPL}$), multipath fading ($L_{\rm MF}$), and building entry loss ($L_{\rm BEL}$). (b) Photograph of the experimental receiving setup. (c) Spectra of the received FM broadcast signal measured using the loop antenna and the RAFEF. For the RAFEF measurement, the broadcast carrier frequency was 93.5~MHz, and a 93.3~MHz local-oscillator (LO) field was applied, producing a difference-frequency signal centred at 200~kHz. For the loop-antenna measurement, no LO field was applied. (d)--(f) Measured signal-to-noise ratio (SNR) as a function of modulation rate at fixed frequency deviations of 15, 35, and 75~kHz, respectively. (g)--(i) Measured SNR as a function of frequency deviation at fixed modulation rates of 1, 7, and 15~kHz, respectively. Red curves show the atomic-receiver results, whereas blue curves show the direct spectrum-analyser reference obtained by connecting the FM signal source directly to the spectrum analyser. For (d)--(i), the FM signal was generated at 92.6~MHz with an output power of -40~dBm and mixed with a 92.8~MHz LO field.}
       \label{fig5-}
    \end{figure*}

The tunable narrowband response established above provides a frequency-selective receiving function that differs from the broadband response of the conventional loop antenna. To directly compare this function, three independently modulated signals at \(f_1=85.3~\mathrm{MHz}\), \(f_2=87.65~\mathrm{MHz}\) and \(f_3=90~\mathrm{MHz}\) were combined and transmitted simultaneously to the two receivers (Fig.~\ref{fig4-}).

With the RAFEF tuned to \(f_1\), the data carried by the \(85.3~\mathrm{MHz}\) channel are preferentially recovered, while the off-resonant components are strongly reduced in the detected waveform. In contrast, the broadband loop antenna receives a superposition of the three incident signals and does not directly recover the target channel without additional channel-selection processing. The comparison demonstrates a frequency-selective receiving advantage of the RAFEF, whose tunable narrowband response provides frequency discrimination directly at the receiving-feed stage, whereas the broadband loop antenna receives the simultaneously incident channels as a superposition.

    \begin{figure*}
        \centering
        \includegraphics[width=1\linewidth]{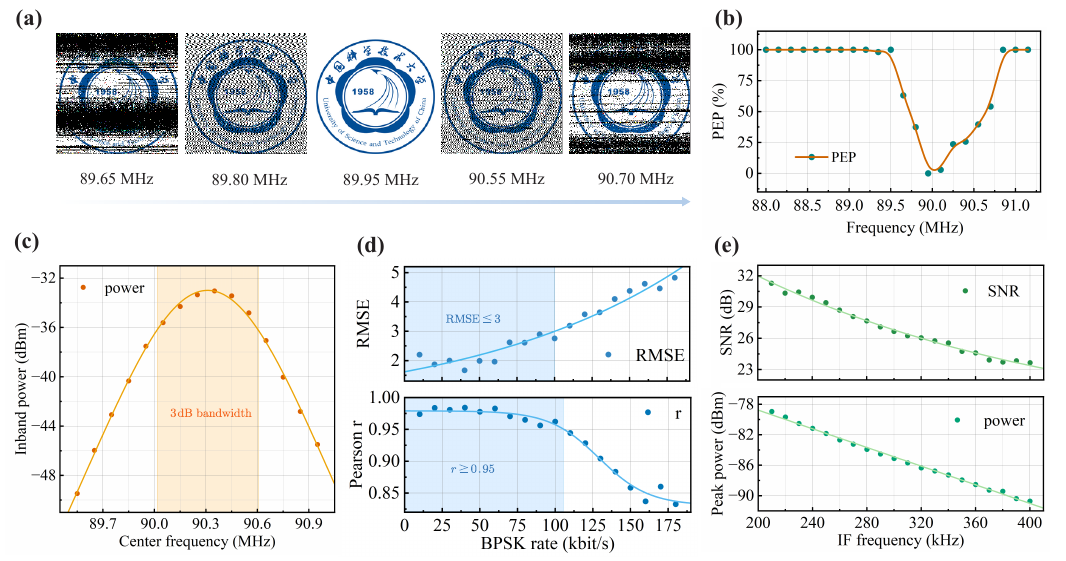}
        \caption{\textbf{Carrier-frequency selectivity and BPSK-modulated signal response of the RAFEF.} (a) Reconstructed images obtained from a BPSK-modulated signal at different carrier frequencies across the RAFEF receiving resonance. (b) Pixel error percentage (PEP) of the reconstructed image as a function of carrier frequency. (c) Received in-band power of the BPSK signal as a function of carrier frequency. The shaded region indicates the $3$-dB carrier-frequency bandwidth. (d) Root-mean-square error (RMSE) and Pearson correlation coefficient $r$ between the transmitted and recovered BPSK signals as functions of BPSK modulation rate. The shaded region indicates the range satisfying RMSE $\leq 3$ and $r\geq0.95$. (e) Received SNR (top) and peak power (bottom) as functions of IF frequency.}\label{fig6-}
    \end{figure*}


\vspace{0.5\baselineskip}
\noindent\textbf{FM-signal reception and over-the-air validation.}
We next evaluated whether the receiving enhancement of the RAFEF is compatible with analogue frequency-modulated (FM) signals, progressing from controlled laboratory characterization to reception of an over-the-air broadcast. The RAFEF-based receiver was first characterized using laboratory-generated FM signals with controlled modulation parameters to determine an appropriate operating condition for FM reception. The optimized receiving configuration was subsequently used to detect a \(93.5~\mathrm{MHz}\) broadcast signal transmitted from a station located approximately \(10~\mathrm{km}\) from the receiver. 

For the controlled measurements, an FM signal at \(92.6~\mathrm{MHz}\) with an output power of \(-40~\mathrm{dBm}\) was applied together with a \(92.8~\mathrm{MHz}\) local-oscillator (LO) field, producing an intermediate-frequency (IF) response centred at \(200~\mathrm{kHz}\). Figures~\ref{fig5-}(d)--(f) show the measured signal-to-noise ratio (SNR) as a function of modulation rate at fixed frequency deviations of \(15\), \(35\), and \(75~\mathrm{kHz}\), respectively. At a frequency deviation of \(15~\mathrm{kHz}\), the SNR measured with the RAFEF-based atomic receiver decreases from approximately \(32\) to \(27~\mathrm{dB}\) as the modulation rate increases from \(1\) to \(15~\mathrm{kHz}\). The dependence becomes weaker at the larger frequency deviations investigated.

Figures~\ref{fig5-}(g)--(i) show the complementary dependence on frequency deviation at fixed modulation rates of \(1\), \(7\), and \(15~\mathrm{kHz}\), respectively. For all three modulation rates, increasing the frequency deviation from \(15\) to \(75~\mathrm{kHz}\) reduces the measured RAFEF SNR by approximately \(5{-}7~\mathrm{dB}\). The red curves in Figs.~\ref{fig5-}(d)--(i) show the RAFEF-based atomic-receiver measurements, whereas the blue curves show the direct spectrum-analyser reference obtained by connecting the FM signal source directly to the spectrum analyser. The two measurement configurations exhibit similar dependencies on modulation rate and frequency deviation. Because they involve different receiving and detection paths, their absolute SNR values are not directly compared; instead, the direct reference is used to assess the parameter-dependent FM response of the RAFEF-based receiver.

Following the laboratory characterization, the RAFEF was adjusted for reception of the real FM broadcast. Figure~\ref{fig5-}(a) illustrates the outdoor reception scenario, in which a \(93.5~\mathrm{MHz}\) broadcast signal was transmitted from a station located approximately \(10~\mathrm{km}\) from the receiver. The corresponding RAFEF-based receiving system and experimental scene are shown in Fig.~\ref{fig5-}(b). For this measurement, the RAFEF was tuned to the broadcast carrier and an LO field at \(93.3~\mathrm{MHz}\) was applied, producing an atomic heterodyne response centred near \(200~\mathrm{kHz}\).

Figure~\ref{fig5-}(c) compares the received spectra obtained using the RAFEF and the conventional metallic loop antenna. A distinct \(200~\mathrm{kHz}\) heterodyne response is observed with the RAFEF-based receiver, yielding an SNR of \(13.60~\mathrm{dB}\). The conventional loop antenna yields an SNR of \(11.07~\mathrm{dB}\), corresponding to a \(2.53~\mathrm{dB}\) lower received SNR than that obtained with the RAFEF. The detection of the \(93.5~\mathrm{MHz}\) broadcast from a transmitter located approximately \(10~\mathrm{km}\) away demonstrates the operation of the RAFEF beyond laboratory-generated signals, while the comparison with the loop antenna shows a higher received SNR for the RAFEF under the present outdoor receiving conditions.


\vspace{0.5\baselineskip}
\noindent\textbf{Carrier-frequency response and BPSK signal recovery.}
We next examined how the narrowband carrier response of the RAFEF is transferred to information-bearing signals and how the receiver responds to changes in modulation rate and IF frequency. In the first experiment, a BPSK-modulated image was transmitted while the carrier frequency was varied across the RAFEF resonance. As shown in Fig.~\ref{fig6-}(a), the reconstructed image exhibits its lowest error when the carrier lies within the RAFEF receiving band, whereas detuning the carrier progressively degrades the reconstructed image. The corresponding pixel error percentage (PEP) is shown in Fig.~\ref{fig6-}(b). The PEP remains low within the receiving band and increases markedly as the carrier is detuned from resonance. These measurements demonstrate that the frequency-selective response of the RAFEF is retained at the information level, enabling reliable BPSK image reception within a finite carrier-frequency range. Details of the BPSK modulation, synchronization and reconstruction procedures are provided in Supplementary Note~4.

The carrier-frequency response was independently quantified using the received in-band power of a BPSK signal. Figure~\ref{fig6-}(c) shows the received in-band power as the BPSK carrier frequency is swept across the RAFEF resonance. The response reaches a maximum near 90.3~MHz and decreases on either side of the resonance, yielding a $3$-dB carrier-frequency bandwidth of nearly $600$~kHz. This measurement defines the RF-frequency range over which a signal can be efficiently received without retuning the RAFEF.

We further evaluated the BPSK signal-response performance as a function of modulation rate. As shown in Fig.~\ref{fig6-}(d), the recovered signal maintains RMSE $\leq 3$ and a Pearson correlation coefficient $r\geq0.95$ up to approximately 100~kbit~s$^{-1}$. At higher modulation rates, the RMSE increases and the correlation decreases, indicating a progressive degradation in waveform recovery. The frequency dependence of the heterodyne response was independently characterized by varying the IF frequency (Fig.~\ref{fig6-}(e)). Both the received SNR and peak power decrease as the IF frequency is increased from 200 to 400~kHz, revealing the finite IF response of the RAFEF.

These measurements show that the RAFEF provides a frequency-selective carrier response while retaining the ability to recover modulated signals over a finite range of modulation rates and IF frequencies. This combination of carrier selectivity and information recovery provides the basis for the frequency-selective multi-channel reception demonstrated below.

    \begin{figure*}
        \centering
        \includegraphics[width=1\linewidth]{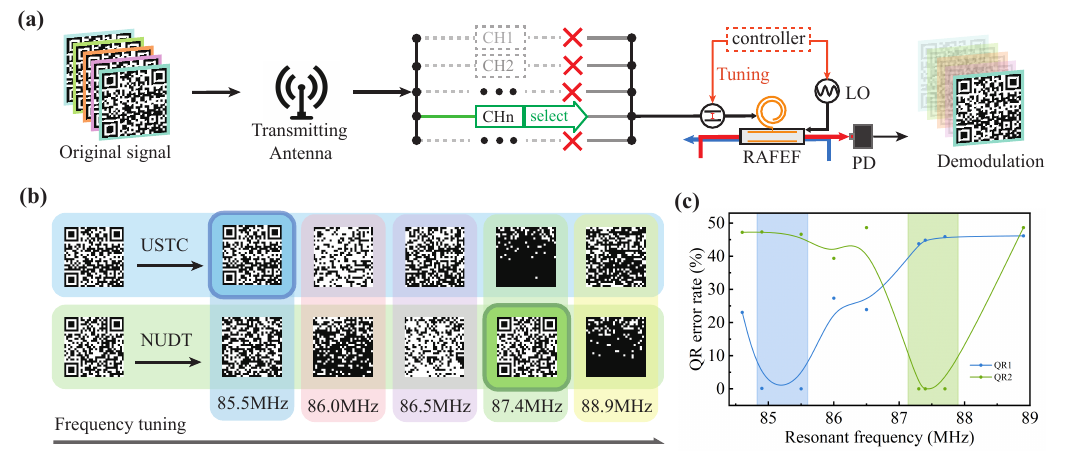}
       \caption{\textbf{Frequency-selective communication using the tunable RAFEF receiver.} (a) Principle of frequency-selective reception. QR-code information is encoded onto different RF channels and transmitted through antennas. By tuning its resonance frequency, the RAFEF selectively converts the target channel into an optical signal while suppressing off-resonant channels. The optical output is detected and subsequently demodulated to recover the selected QR code. (b) Experimental demonstration of selective QR-code recovery. The upper row shows the original QR1 pattern encoding ``USTC'' and the corresponding reconstructed patterns obtained at different RAFEF tuning frequencies. The lower row shows the original QR2 pattern encoding ``NUDT'' and its reconstructed patterns. Each QR code is faithfully recovered when the RAFEF is tuned to its corresponding channel, whereas the off-resonant QR code cannot be correctly demodulated. (c) QR-code error rate as a function of the RAFEF tuning frequency. Blue and green symbols represent QR1 and QR2, respectively. The two error-rate minima occur near the corresponding channel frequencies, demonstrating tunable frequency-selective reception and off-channel suppression.}\label{fig7-}
    \end{figure*}

\vspace{0.5\baselineskip}
\noindent\textbf{Frequency-selective multi-channel reception.}
We finally exploited the intrinsic narrowband response and frequency tunability of the RAFEF for selective reception across different RF channels. As illustrated in Fig.~\ref{fig7-}(a), independent BPSK-modulated signals were transmitted separately at different carrier frequencies and received by the RAFEF-based atomic receiver. Rather than separating the channels solely through post-detection filtering, the RAFEF resonance was tuned to preferentially receive the target channel while reducing the response to off-resonant channels. The RAFEF thus provides frequency discrimination directly at the receiving-feed stage.

Figure~\ref{fig7-}(b) demonstrates this frequency-selective response at the information-recovery level using two independently evaluated BPSK channels carrying independent QR codes. QR1 ("USTC") and QR2 ("NUDT") were transmitted at carrier frequencies of \(85.5~\mathrm{MHz}\) and \(87.5~\mathrm{MHz}\), respectively, with the corresponding LO frequencies chosen to maintain a common IF of \(200~\mathrm{kHz}\). When the RAFEF is tuned near the QR1 carrier, the reconstruction error of QR1 is minimized while that of QR2 increases. Tuning the RAFEF near the QR2 carrier reverses this response. The experimental configuration and signal parameters used for frequency-selective QR-code reception are described in Supplementary Note~3.

The frequency-selective reception was further quantified by measuring the QR-code error rate as a function of RAFEF tuning frequency (Fig.~\ref{fig7-}(c)). The error rate for QR1 reaches a minimum near its \(85.5~\mathrm{MHz}\) carrier frequency, whereas that for QR2 reaches a separate minimum near \(87.5~\mathrm{MHz}\). For each channel, the error rate increases as the RAFEF is detuned from the corresponding carrier, quantitatively confirming the frequency-selective information recovery observed in Fig.~\ref{fig7-}(b). The experiment therefore extends the frequency-selective advantage established against the broadband loop antenna in Fig.~\ref{fig4-} from a fixed-channel benchmark to tunable information-level channel selection.

\section*{Discussion}

The central advance of this work is the realization of a RAFEF that integrates electromagnetic field coupling with localized atomic field-to-optical transduction. Rydberg atoms alone provide a sensitive local-field transduction mechanism, but their response depends on how effectively the incident electromagnetic field is coupled into the finite optically interrogated atomic volume. The RAFEF addresses this receiving-interface problem by combining incident-field capture, resonant field concentration and atomic-volume overlap within a unified receiving architecture. It should therefore be viewed not simply as a resonant field-enhancement structure, but as an electromagnetic--atomic receiving feed in which the coupling and transduction processes are jointly considered.

The distinction from a conventional electrically small antenna lies in the mechanism by which the incident field is coupled to and transduced by the sensing element, rather than in circumventing the fundamental bounds of passive electrically small electromagnetic structures. A conventional antenna converts the incident electromagnetic field into an electrical response that is extracted through an antenna port. In the RAFEF-based receiver, the incident field is instead concentrated within a deeply subwavelength atomic interaction volume and locally transduced through Rydberg-state dynamics into an optical response. The measurements presented here show that this architecture can provide lower source-power requirements for a given measured SNR under the investigated benchmark conditions and can additionally provide frequency discrimination directly at the receiving-feed stage.

The receiving-performance comparison further shows why peak enhancement alone is insufficient to characterize a resonant atomic receiving feed. Increasing resonant field concentration can improve sensitivity, but this improvement must be considered together with the instantaneous receiving bandwidth and the electrical size of the complete receiving architecture. The receiving-capability metric \(\Phi\) introduced here provides a system-level measure of this trade-off by combining receiving enhancement, instantaneous 3-dB receiving bandwidth and normalized three-dimensional volume. Under this definition, both RAFEF configurations yield higher \(\Phi\) values than the conventional loop-antenna benchmark. The comparison therefore shows that the measured enhancement of the RAFEF remains favourable when the bandwidth and volume terms included in the metric are taken into account, rather than reflecting peak enhancement alone. More generally, these results indicate that the performance of an atomic receiving feed should be considered through the joint optimization of electromagnetic coupling, atomic-volume overlap, instantaneous bandwidth and physical size, rather than by maximizing local field enhancement in isolation.

The intrinsically narrowband response of the RAFEF also plays a dual role. Spectral confinement restricts the carrier-frequency interval that can be received without retuning, but at the same time provides frequency discrimination before subsequent signal processing. The carrier-frequency measurements show that this selectivity is retained at the information-recovery level, while the multi-channel experiments demonstrate that tuning the RAFEF resonance enables one target channel to be preferentially recovered in the presence of other simultaneously incident signals. In this sense, the narrowband response is not solely a bandwidth limitation. When combined with frequency tunability, it becomes a functional property of the receiving feed, providing front-end frequency selection that is absent from the broadband loop antenna used for comparison. The BPSK and FM measurements further show that this receiving enhancement and selectivity remain compatible with modulated information signals.

The measurements of different RAFEF structures further show that the receiving architecture can be transferred across different VHF operating frequencies through systematic modification of the logarithmic-spiral geometry. Increasing the logarithmic-spiral turn number shifts the resonance towards lower frequencies while retaining receiving enhancement. Because each point corresponds to a different fabricated structure, the observed frequency dependence should be interpreted as a design trend across RAFEF geometries rather than as the broadband response or a universal scaling law of a single device. Extending this approach to lower frequencies will require further co-optimization of electrical size, structural loss and field overlap with the atomic interaction volume.

Several limitations remain. The present RAFEF relies on mechanical adjustment of the resonant structure, which limits tuning speed and may affect long-term frequency stability. Electronically reconfigurable implementations could enable faster and closed-loop frequency control. Further improvement will also require co-optimization of the electromagnetic mode distribution, its overlap with the finite atomic interaction volume, the atomic transduction process and the optical and electronic readout bandwidth. Such co-design will be important for extending modulation and IF response while preserving the receiving enhancement and frequency selectivity of the feed. Overall, these results establish the RAFEF as an electromagnetic--atomic receiving-feed architecture that exploits localized electric-field transduction to provide a route towards compact, tunable and frequency-selective long-wavelength signal reception beyond conventional electrically small antenna feeds.

\section*{Methods}

\textbf{Experimental setup of the RAFEF receiving architecture.}
The experimental configuration of the Rydberg atomic field-enhanced feed (RAFEF)-based receiver is illustrated in Fig.~\ref{fig1}. The system consists of three integrated components: a tunable RAFEF receiving feed, a Rydberg atomic transduction module and an optical heterodyne detection subsystem. The RAFEF is positioned around the atomic vapor cell to couple incident radio-frequency (RF) fields into the atomic interaction region, where the RF-induced atomic response is converted into an optical signal.

The receiving process begins with the coupling of the incident RF signal into the RAFEF. The resonant electromagnetic mode generated by the feed interacts with the Rydberg ensemble and modifies the optical response of the atoms. The RF-dependent optical modulation is subsequently extracted through heterodyne detection, in which the transmitted probe field is mixed with a reference local oscillator (LO) field to generate an intermediate-frequency (IF) signal.

The atomic transduction module is based on a ladder-type Rydberg excitation scheme in a thermal \(^{87}\mathrm{Rb}\) vapor cell. A probe laser and a coupling laser establish electromagnetically induced transparency (EIT) between the ground state and the selected Rydberg state. The incident RF field couples adjacent Rydberg levels and induces an RF-dependent modification of the optical transmission spectrum. The resulting optical response is detected by a photodetector and subsequently processed for spectrum analysis, waveform reconstruction and communication demodulation.

The RF signal and LO field are generated independently and applied to the RAFEF-based receiver. By controlling the frequency difference between the RF signal and the LO field, the received electromagnetic information is converted into an IF optical response. The same receiving architecture is used for all measurements presented in this work, including sensitivity characterization, frequency tuning, communication signal reception and frequency-selective channel recovery, while the atomic and optical configurations remain unchanged.

\textbf{Design and characterization of the RAFEF.}
The RAFEF consists of a tunable resonant electromagnetic structure integrated with the atomic vapor cell. The geometry of the feed was optimized to achieve the desired operating frequency while maintaining effective electromagnetic coupling between the resonant mode and the atomic interrogation region. The electromagnetic structure, vapor cell position and optical access configuration were considered simultaneously during the design process.

Full-wave electromagnetic simulations were performed to characterize the resonance response and electric-field distribution of the RAFEF. The simulated frequency response was used to determine the resonance characteristics of the feed, while the calculated electric-field distribution was analysed to evaluate the spatial overlap between the resonant electromagnetic mode and the atomic sensing region.

The operating frequency of the RAFEF was tuned by modifying the effective electromagnetic boundary condition of the resonant structure. This tuning approach enables the same receiving architecture to operate at different low-frequency bands while preserving the atomic transduction scheme and optical readout configuration.

The fabricated RAFEF was experimentally characterized through frequency-domain measurements under RF excitation. The measured resonance response and tuning range were compared with the electromagnetic simulations. The atomic response obtained with different RAFEF configurations was subsequently measured to verify the coupling between the engineered electromagnetic mode and the Rydberg atomic ensemble.

The electromagnetic response and physical dimensions of the complete RAFEF architecture were further characterized for evaluation of the receiving performance. The considered structure includes the electromagnetic feed, atomic vapor cell and associated components required for operation, ensuring that the characterization represents the complete atomic receiving feed rather than an isolated resonant element.

\textbf{Rydberg atomic transduction and optical readout.}
The RF-to-optical transduction in the RAFEF-based receiver is realized through the interaction between the electromagnetic field delivered by the RAFEF and the Rydberg atomic ensemble. The atomic transduction module employs a ladder-type Rydberg excitation scheme in a thermal \(^{87}\mathrm{Rb}\) vapor cell. A probe laser and a coupling laser are used to establish electromagnetically induced transparency (EIT) between the ground state and the selected Rydberg state, providing an optical interface for detecting RF-induced atomic responses.

The RF field coupled into the atomic interaction region interacts with the Rydberg transition and modifies the atomic coherence through the non-resonant AC Stark response. The RF-induced modification of the Rydberg energy levels changes the optical transmission spectrum of the EIT system, resulting in an RF-dependent optical modulation. This process provides the transduction mechanism that converts the electromagnetic signal captured by the RAFEF into an optically detectable response.

The optical readout is performed using a heterodyne detection scheme. The transmitted probe field from the atomic ensemble is combined with a reference local oscillator (LO) field, generating a beat signal at the frequency difference between the RF signal and the LO field. The intermediate-frequency (IF) signal is determined by

\begin{equation}
f_{\mathrm{IF}}=\left|f_{\mathrm{RF}}-f_{\mathrm{LO}}\right|,
\end{equation}

where \(f_{\mathrm{RF}}\) and \(f_{\mathrm{LO}}\) represent the frequencies of the received RF signal and the LO field, respectively. The generated optical IF signal is detected by a photodetector and subsequently digitized for spectrum analysis, waveform reconstruction and communication demodulation.

For communication experiments, the LO frequency is adjusted according to the target RF carrier frequency while maintaining a fixed IF offset for signal processing. This heterodyne configuration enables the recovery of both digital and analogue modulation formats, including BPSK and frequency-modulated (FM) signals, using the same RAFEF-based receiving architecture.

During the characterization of the RAFEF receiving performance, the atomic excitation and optical detection parameters are maintained unchanged. Therefore, variations in the measured RF response primarily originate from the electromagnetic coupling characteristics of the RAFEF rather than modifications of the atomic transduction process. The atomic module consequently functions as a stable RF-to-optical conversion interface, allowing the receiving capability of different RAFEF configurations to be directly compared.

\textbf{Receiving performance characterization.}
The receiving performance of the RAFEF-based atomic receiver was characterized through measurements of sensitivity enhancement, tunable signal reception, communication recovery and frequency-selective channel discrimination. The same experimental configuration and atomic readout conditions were maintained throughout all measurements, while the electromagnetic response of the RAFEF was adjusted according to the target operating frequency.

The sensitivity enhancement provided by the RAFEF was characterized by measuring the minimum detectable RF signal under controlled excitation conditions. An RF signal and a local oscillator (LO) field were simultaneously applied to the receiving feed, and the resulting optical heterodyne response was recorded as a function of the incident RF power.

The signal-to-noise ratio (SNR) was extracted from the measured intermediate-frequency spectrum as

\begin{equation}
\mathrm{SNR}
=
10\log_{10}
\left(
\frac{P_{\mathrm{signal}}}
{P_{\mathrm{noise}}}
\right),
\end{equation}

where \(P_{\mathrm{signal}}\) and \(P_{\mathrm{noise}}\) represent the signal and noise powers obtained from the measured spectrum.

The RAFEF and loop antenna were characterized using their respective receiving and detection configurations. Neither receiver employed external low-noise amplification. The RAFEF and loop antenna employ different transduction and detection mechanisms: the RAFEF converts the locally enhanced RF electric field through the Rydberg atomic response into an optical signal, whereas the loop antenna is read out through its electrical receiving chain. The comparison was therefore performed at the system level by relating the measured output SNR to the applied signal-generator output power using the respective receiver-specific detection chains.

The frequency-tunable receiving capability of the RAFEF was characterized by adjusting the resonance frequency of the feed and recording the corresponding atomic heterodyne response. The carrier-frequency response was obtained by sweeping the RF signal frequency while maintaining the same atomic and optical configurations.

The communication capability was evaluated using both digital and analogue modulated signals. For BPSK measurements, digitally modulated RF signals with different symbol rates were transmitted to the RAFEF-based receiver. The optical heterodyne output was digitized and processed for signal reconstruction and demodulation. The recovered waveform was compared with the transmitted sequence to evaluate the signal fidelity.

For FM measurements, frequency-modulated RF signals with controlled modulation parameters were generated and received using the same RAFEF-based architecture. The recovered intermediate-frequency signal was demodulated to evaluate the preservation of analogue modulation information. The over-the-air FM broadcast experiment was performed using the same receiving configuration without modification of the atomic transduction module.

The frequency-selective receiving capability of the RAFEF was evaluated using multiple RF channels carrying independent information. Multiple carrier frequencies were transmitted while the resonance frequency of the RAFEF was tuned to different target channels.

The received information was recovered directly from the atomic heterodyne response without additional RF filtering before the receiving feed. The channel discrimination performance was evaluated by analysing the reconstruction accuracy as a function of the RAFEF resonance frequency.

This measurement characterizes the ability of the RAFEF to selectively convert spectrally matched RF signals while suppressing off-resonant channels through the intrinsic frequency response of the receiving feed.

\textbf{Evaluation of the RAFEF receiving capability \(\Phi_{\mathrm{RAFEF}}\).}

The receiving capability of the RAFEF was evaluated from experimentally extracted parameters describing the sensitivity enhancement, effective receiving bandwidth and normalized three-dimensional volume of the complete atomic receiving architecture. These parameters were obtained independently from electromagnetic characterization and RF receiving measurements.

The sensitivity enhancement was determined from RF power measurements by comparing the minimum detectable signal level of the RAFEF-enabled receiver with that of the reference configuration without electromagnetic field enhancement. The comparison was performed under identical RF excitation, atomic transduction and optical detection conditions to ensure that the extracted sensitivity improvement originated from the receiving feed.

The receiving bandwidth was defined as the fixed-state carrier-frequency
interval over which the received power remained within 3~dB of its maximum value.

The effective physical volume was determined from the three-dimensional dimensions of the complete RAFEF architecture. The volume estimation includes the electromagnetic receiving structure, the atomic vapor cell and the associated components required for RF coupling and optical interrogation.This definition represents the effective enclosing volume of the complete atomic receiving feed rather than the volume of the resonant element alone.

The experimentally obtained sensitivity enhancement, receiving bandwidth and normalized physical volume were subsequently used to evaluate the system-level receiving capability of the RAFEF. This evaluation enables direct comparison between different atomic receiving architectures by simultaneously considering receiving performance and physical footprint.

{\color{black}
\section*{DATA AVAILABILITY}
        All experimental data used in this study are available from the corresponding author upon request.

\section*{CODE AVAILABILITY}
 Custom codes used to produce the results presented in this paper are available from the corresponding authors on request.
}

\section*{Acknowledgements}
       We acknowledge funding from the National Natural Science Foundation of China (Grant Nos. T2495253, 62435018), the National Key R and D Program of China (Grant No. 2022YFA1404002).

\section*{Author contributions statement}
       D.-S.D. conceived the idea and supported this research. J.D.N., Q.Q.Z. and H.Y.F. conducted the physical experiments. Q.Q.Z. conducted the theoretical calculations. The manuscript was written by Q.Q.Z. and J.D.N.. All authors contributed to discussions regarding the results and the analysis contained in the manuscript.

\section*{Competing interests}
    The authors declare no conflict of interest.


\bibliography{ref}

\end{document}


\title{Supplementary Information : Rydberg atomic field-enhanced feed beyond electrically small antennas}

\author{Jia-Dou Nan$^{1,2,6,\textcolor{blue}{\star}}$}
\author{Qian-Qian Zhang$^{1,2,3,\textcolor{blue}{\star}}$}
\author{Hao-Yu Fan$^{3,\textcolor{blue}{\star}}$}
\author{Jin-Hang Fang$^{4}$}
\author{Luan-Yi Wei$^{4}$}
\author{Bang Liu$^{1,2,4}$}
\author{Kai Wang$^{3,\textcolor{blue}{\S}}$}
\author{Li-Hua Zhang$^{1,2,4,\textcolor{blue}{\ddagger}}$}
\author{Qing-Li Ma$^{3}$}
\author{Bo Du$^{5,\textcolor{blue}{\star\star}}$}
\author{Dong-Sheng Ding$^{1,2,4,6,\textcolor{blue}{\dagger}}$}
\author{Bao-Sen Shi$^{1,2,6}$}

\affiliation{$^1$Key Laboratory of Quantum Information, University of Science and Technology of China, Hefei, Anhui 230026, China.}
\affiliation{$^2$Anhui Province Key Laboratory of Quantum Network, University of Science and Technology of China, Hefei, Anhui 230026, China.}
\affiliation{$^3$College of Electronic Engineering, National University of Defense Technology, Hefei, Anhui 230037, China}
\affiliation{$^4$Davinci Quantum Technology Co., Ltd, Hefei, Anhui 230026, China}
\affiliation{$^5$Qianyuan Laboratory, Hangzhou, Zhejiang 310024, China.}
\affiliation{$^6$Hefei National Laboratory, University of Science and Technology of China, Hefei, 230088, China.}
\date{\today}

\symbolfootnote[1]{J.D.N., Q.Q.Z. and H.Y.F. contribute equally to this work.}
\symbolfootnote[4]{wkhy@ustc.edu.cn}
\symbolfootnote[3]{zlhphys@ustc.edu.cn}
\symbolfootnote[6]{dubo@zju.edu.cn}
\symbolfootnote[2]{dds@ustc.edu.cn}
\date{\today}

\maketitle

\color{black}

\section{Supplementary Notes}

\subsection*{Supplementary Note 1. RF-field enhancement and uniformity}

\begin{figure}[H]
        \centering
        \includegraphics[width=1\linewidth]{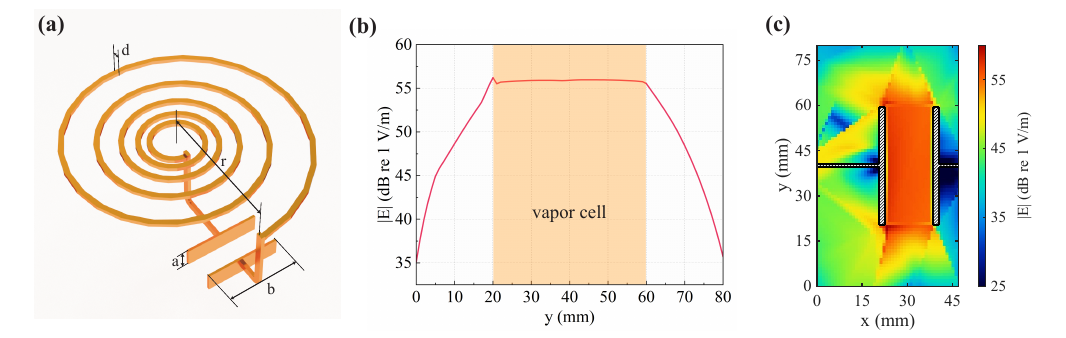}
       \caption{\textbf{RAFEF geometry and field distribution.} (a) Structural dimensions of the resonant enhancement structure. The parameters $a$, $b$, $d$, and $r$ are 10, 40, 3, and 74~mm, respectively. (b) Simulated electric-field magnitude along the $y$ direction. The field magnitude is expressed in decibels referenced to 1~V~m$^{-1}$. The orange shaded region indicates the plate position, where the field enhancement is nearly uniform. (c) Two-dimensional electric-field distribution in the $x$--$y$ plane. The black hatched regions denote the plates, and the colour scale represents the local electric-field magnitude.}
        \label{figS1}
    \end{figure}

Supplementary Fig.~\ref{figS1} characterizes the geometry and simulated electric-field distribution of the resonant enhancement structure. As shown in Supplementary Fig.~\ref{figS1}(a), the logarithmic spiral is coupled to a parallel-plate interaction region with structural parameters \(a=10~\mathrm{mm}\), \(b=40~\mathrm{mm}\), \(d=3~\mathrm{mm}\), and \(r=74~\mathrm{mm}\). The spiral section collects and concentrates the incident VHF field, whereas the parallel plates form an extended interaction region that overlaps with the Rydberg vapor cell and the optical beams.

The electric-field profile along the longitudinal \(y\) direction is shown in Supplementary Fig.~\ref{figS1}(b). The field magnitude increases rapidly on entering the parallel-plate region and reaches a plateau of approximately \(55{-}56~\mathrm{dB~re~1~V\,m^{-1}}\) over \(y\approx20{-}60~\mathrm{mm}\). The variation within this region is less than approximately \(1~\mathrm{dB}\), indicating that the enhanced field is nearly uniform along most of the plate length. Outside the interaction region, the field magnitude decreases by approximately \(20~\mathrm{dB}\), confirming that the resonant structure confines the enhanced field predominantly to the intended atomic sensing volume.

The two-dimensional distribution in Supplementary Fig.~\ref{figS1}(c) further shows that the enhanced field forms an extended high-field region between the parallel plates rather than a strongly localized point hotspot. The largest field gradients occur near the plate ends and conducting boundaries, while the central region between the plates remains comparatively uniform. This spatial distribution is advantageous for ensemble-based Rydberg detection because a larger fraction of the optically interrogated atoms experiences a similar RF-field amplitude.

Spatial uniformity is particularly important in the present receiver because the measured optical transmission represents an average response over the finite laser--atom interaction volume. A highly localized field would interact with only a small fraction of the atomic ensemble and generate spatially varying AC Stark shifts, resulting in inhomogeneous broadening and a reduced effective optical response. In contrast, the extended and nearly uniform field distribution increases the spatial overlap between the resonant electromagnetic mode and the Rydberg ensemble, thereby allowing more atoms to contribute coherently to the detected heterodyne signal.

These simulations therefore indicate that the resonant structure provides not only local field concentration but also an interaction region suitable for ensemble-averaged atomic optical readout. The resulting combination of field enhancement and spatial uniformity supports the experimentally observed improvement in weak-signal detection while reducing the adverse effects associated with strongly nonuniform atom--field coupling.


\FloatBarrier
\subsection*{Supplementary Note 2. Design rationale of the logarithmic-spiral RAFEF}

\begin{table}[htbp]
\centering
\caption{Geometrical parameters of the logarithmic spirals used in the RAFEF simulations and fabricated devices. Two nominally identical RAFEF2 devices, denoted RAFEF2-A and RAFEF2-B, were fabricated using the same parameters listed here. Their measured resonance frequencies differ slightly, most likely owing to fabrication tolerances. The additional five RAFEF devices share the same initial radius and growth coefficient and differ only in the number of turns.}
\label{tab:rafef_parameters}

\begin{tabular}{cccc}
\toprule
Device
& Initial radius ($r_0$)
& Number of turns ($N$)
& Growth coefficient ($\alpha$) \\
\midrule

\multicolumn{4}{l}{\textit{Devices used in the main experiments}} \\
RAFEF~1 & $10~\mathrm{mm}$ & $5$ & $0.062$ \\
RAFEF~2 (A and B) & $11~\mathrm{mm}$ & $5$ & $0.060$ \\

\addlinespace
\multicolumn{4}{l}{\textit{Additional devices fabricated for the turn-number study}} \\
RAFEF ($N=5$) & $13~\mathrm{mm}$ & $5$ & $0.040$ \\
RAFEF ($N=6$) & $13~\mathrm{mm}$ & $6$ & $0.040$ \\
RAFEF ($N=7$) & $13~\mathrm{mm}$ & $7$ & $0.040$ \\
RAFEF ($N=8$) & $13~\mathrm{mm}$ & $8$ & $0.040$ \\
RAFEF ($N=9$) & $13~\mathrm{mm}$ & $9$ & $0.040$ \\

\bottomrule
\end{tabular}
\end{table}

A Rydberg atomic receiver detects the local RF electric field within the optically interrogated atomic volume. Efficient reception therefore requires the incident free-space field to be collected and subsequently concentrated within this finite sensing region. In the RAFEF, these functions are separated between the spiral conductor and the parallel plates. The spiral acts as an inductive receiving element, whereas the plates provide a localized capacitance and convert the resonantly enhanced voltage into an electric field overlapping with the atomic ensemble. The resulting response can be described approximately by
\begin{equation}
f_0\simeq\frac{1}{2\pi\sqrt{L_{\mathrm{eff}}C_{\mathrm{eff}}}},
\qquad
E_{\mathrm a}\simeq\frac{V_{\mathrm p}}{d},
\end{equation}
where $L_{\mathrm{eff}}$ and $C_{\mathrm{eff}}$ are the effective inductance and capacitance, while $V_{\mathrm p}$ and $d$ denote the RF voltage and separation between the plates, respectively. The spiral--plate architecture thus provides a direct electromagnetic interface that converts the incident field into an enhanced local field within the atomic interaction volume.

The logarithmic spiral is introduced to provide a scalable receiving architecture. Its centreline is described by
\begin{equation}
r(\theta)=r_0e^{\alpha\theta},
\qquad
r(\theta+\Delta\theta)
=e^{\alpha\Delta\theta}r(\theta)
=s\,r(\theta),
\end{equation}
where \(r_0\) is the initial radius, \(\alpha\) is the logarithmic growth coefficient, and \(s=e^{\alpha\Delta\theta}\) is the geometrical scaling factor. Consequently, extending a logarithmic spiral along its angular coordinate is equivalent to scaling its original geometry without changing the structural topology. According to electromagnetic scaling, when the entire structure is uniformly enlarged by a factor \(s\), its characteristic frequency approximately follows
\begin{equation}
f_0^{\prime} \simeq \frac{f_0}{s}.
\end{equation}

\begin{figure}[H]
    \centering
    \includegraphics[width=1\linewidth]{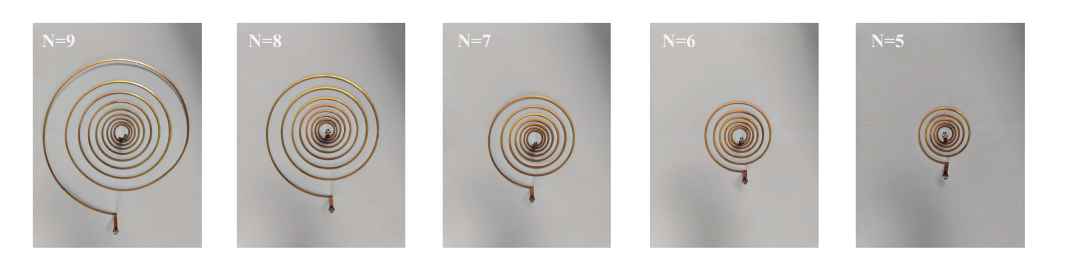}
    \caption{Photographs of five fabricated RAFEF devices with logarithmic-spiral turn numbers ranging from $N=5$ to $N=9$. All RAFEF devices have the same initial radius $r_0=13~\mathrm{mm}$ and growth coefficient \(\alpha=0.04\).}
    \label{figS2}
\end{figure}

Therefore, RAFEFs targeting different frequency bands can retain the same structural topology and normalized geometrical parameters, with their initial dimensions determined through a predictable scaling transformation. The logarithmic-spiral geometry thus facilitates the extension of the RAFEF architecture to other operating frequencies. This scaling relation describes the design transfer between geometrically scaled devices and does not imply ideal frequency-independent operation of a single finite, capacitively loaded RAFEF.

\FloatBarrier
\subsection*{Supplementary Note 3. Frequency-selective communication experiment}

The frequency-selective communication experiment was performed using two Ceyear 1465F-V signal generators to independently generate the signal (SIG) and local-oscillator (LO) fields. The SIG carrier was BPSK-modulated at a symbol rate of 100~ksps, with the RF output power of the signal generator set to -38~dBm. The LO generator was operated in continuous-wave mode with an output power of 5~dBm. The SIG and LO outputs were radiated by a whip antenna and a double-loop antenna, respectively.

Two communication channels were investigated. For channel 1, the SIG and LO frequencies were set to 85.5 and 85.3~MHz, respectively, producing a frequency difference of 200~kHz. The SIG carrier was modulated with a $29\times29$-module QR code encoding ``USTC'' (QR1). The resonance frequency of the RAFEF was tuned across the frequency range of interest, and the resulting optical signal was detected using a photodetector. The photodetector output was acquired by a software-defined radio (SDR) and digitally demodulated to reconstruct the transmitted QR code.

For channel 2, the SIG and LO frequencies were changed to 87.5 and 87.3~MHz, respectively, while maintaining the same 200~kHz frequency offset. A second $29\times29$-module QR code encoding ``NUDT'' (QR2) was transmitted and recovered using the same acquisition and demodulation procedure. The measurements were repeated at different RAFEF tuning frequencies to determine the QR-code error rate as a function of the receiver resonance frequency.

\FloatBarrier
\subsection*{Supplementary Note 4. Signal modulation and demodulation experiments}

Binary phase-shift keying (BPSK) was used to evaluate digital signal transmission through the device. For an input bit $b_k\in\{0,1\}$, the transmitted symbol was defined as $a_k=1-2b_k$. The modulated signal is expressed as
\begin{equation}
s(t)=A\sum_k a_kp(t-kT_s)\cos(2\pi f_ct+\phi_0),
\end{equation}
where $A$ is the signal amplitude, $p(t)$ is the pulse-shaping waveform, $T_s$ is the symbol period and $f_c$ is the carrier frequency. Signals were generated using either a Ceyear 1465F-V vector signal generator or a USRP X310, acquired using an SDRplay RSP1 and demodulated in GNU Radio.

For QR-code transmission, two version-3 QR codes, each containing $29\times29=841$ binary modules, were used as short test payloads. A 32-bit frame header was added before each payload. The hexadecimal frame headers were \texttt{A5C3F03C} for QR1 and \texttt{D39A6C55} for QR2. Each frame contained 873 bits and was zero-padded to 880 bits.

The complete received bitstream was searched for both the original and bit-inverted frame headers to account for the $\pi$-phase ambiguity of BPSK. When an exact header was not detected, candidate positions were identified by correlation with the reference header. Each recovered payload was compared with the corresponding reference QR code, and the candidate with the lowest module error rate was retained. The QR-code module error rate was calculated as $P_{\mathrm{QR}}=(N_{\mathrm{err}}+N_{\mathrm{miss}})/841\times100\%$, where $N_{\mathrm{err}}$ and $N_{\mathrm{miss}}$ are the numbers of erroneous and missing modules, respectively. This reference-assisted analysis reports the best recoverable frame within each acquisition rather than the average error rate of all received frames.

For image transmission, colour images were converted into RGB bitstreams. A 512-bit global synchronisation sequence was added at the beginning of each image. Each pixel was represented by a 64-bit record comprising a 16-bit pixel synchronisation word, two 12-bit coordinates and a 24-bit RGB value. A $200\times200$-pixel image therefore contained $2{,}560{,}512$ bits.

During image reconstruction, both normal and inverted pixel synchronisation words were examined. The coordinate fields were used to place each recovered RGB value at its original position, allowing synchronisation to be re-established after local bit errors. When multiple colour estimates were recovered for the same coordinate, the estimate closest to the reference image was selected. The pixel error rate was defined as $P_{\mathrm{pixel}}=(N_{\mathrm{wrong}}+N_{\mathrm{miss}})/(WH)\times100\%$, where $N_{\mathrm{wrong}}$ is the number of incorrectly reconstructed pixels, $N_{\mathrm{miss}}$ is the number of missing pixels, and $W$ and $H$ are the image dimensions. Missing pixels were included in the total number of erroneous pixels.

Representative image- and QR-code-reconstruction results are shown in Figs.~6 and 7 of the main text. Randomly distributed isolated errors are primarily attributed to noise, whereas spatially correlated stripes and periodic patterns indicate carrier- or timing-synchronisation errors, periodic interference, or bandwidth-dependent signal distortion.

\FloatBarrier
\subsection*{Supplementary Note 5. Receiving bandwidth of the conventional loop antenna}

The frequency response of the conventional loop antenna was measured under a fixed excitation and detection configuration. As shown in Supplementary Fig.~\ref{figS3}, the received power remains within 3~dB of its maximum value from approximately \(88.75\) to \(94.6~\mathrm{MHz}\), corresponding to a 3-dB receiving bandwidth of approximately \(5.9~\mathrm{MHz}\). This measured bandwidth was used to calculate the receiving-capability metric of the loop antenna in the main text.

\begin{figure}[H]
    \centering
    \includegraphics[width=0.35\linewidth]{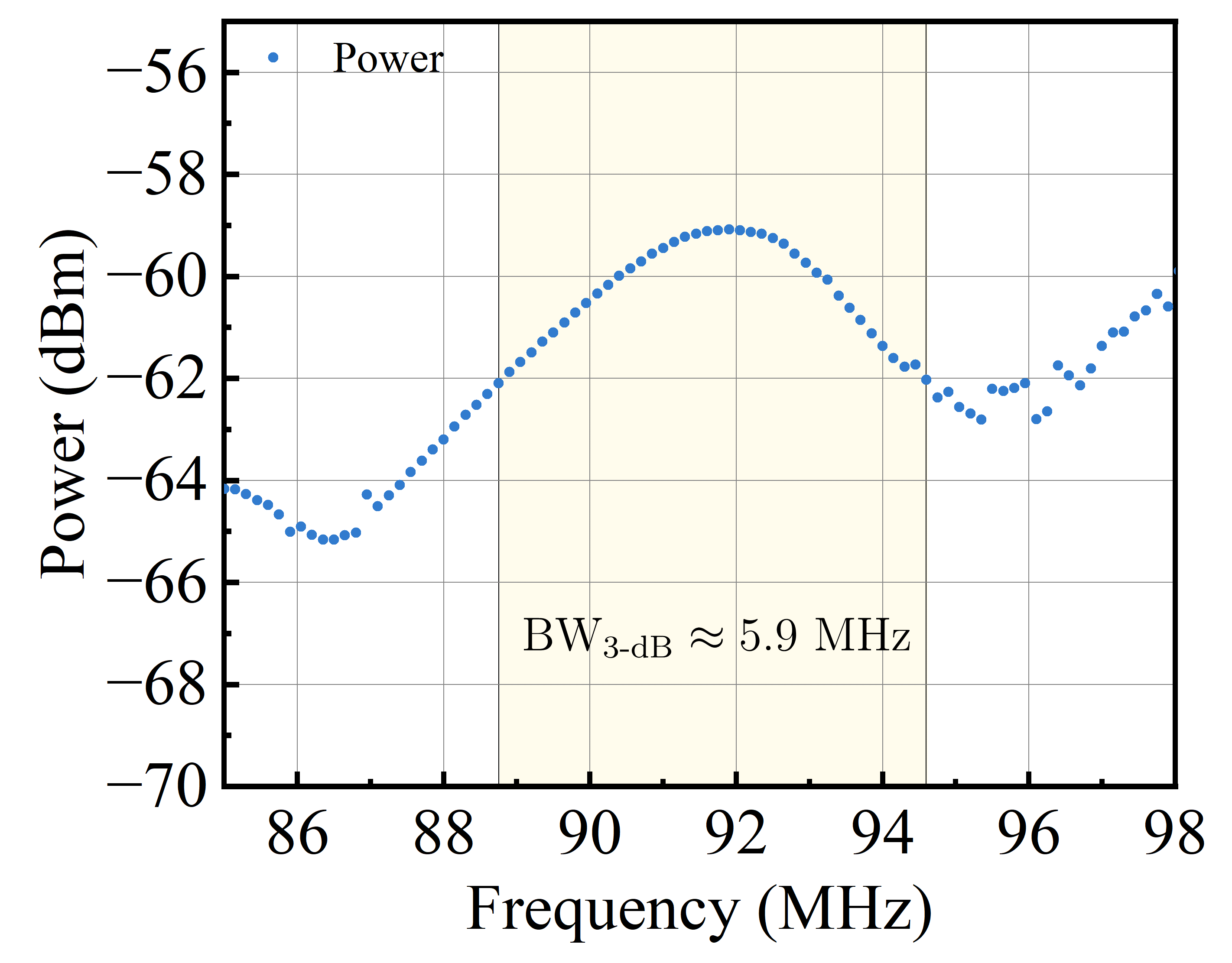}
    \caption{\textbf{Measured loop-antenna bandwidth.}
    Received power as a function of excitation frequency under a fixed measurement configuration. The shaded region indicates the frequency range over which the received power remains within 3~dB of its maximum value. The corresponding 3 dB receiving bandwidth extends from approximately \(88.75\) to \(94.6~\mathrm{MHz}\), giving \(BW_{3\text{-}\mathrm{dB}}\approx5.9~\mathrm{MHz}\).}
    \label{figS3}
\end{figure}

\FloatBarrier
\subsection*{Supplementary Note 6. LNA-assisted loop-antenna response}

In the main text, the RAFEF performance is compared with that of a conventional loop antenna without low-noise amplification, so that the comparison reflects the passive receiving capability of the two front ends under the same measurement conditions. For completeness, we also measured the response of the conventional loop antenna with an additional low-noise amplifier (LNA), as shown in Supplementary Fig.~\ref{figS4}. The received power was recorded as a function of signal-source output power for the loop antenna with and without the LNA. Linear fits to the high-power response give an input-power difference of approximately 12.6~dB at a received power of -115.76~dBm. This result confirms that active amplification improves the readout level of the conventional loop-antenna receiver, while the comparison in the main text focuses on the passive front-end performance.

\begin{figure}[H]
    \centering
    \includegraphics[width=0.35\linewidth]{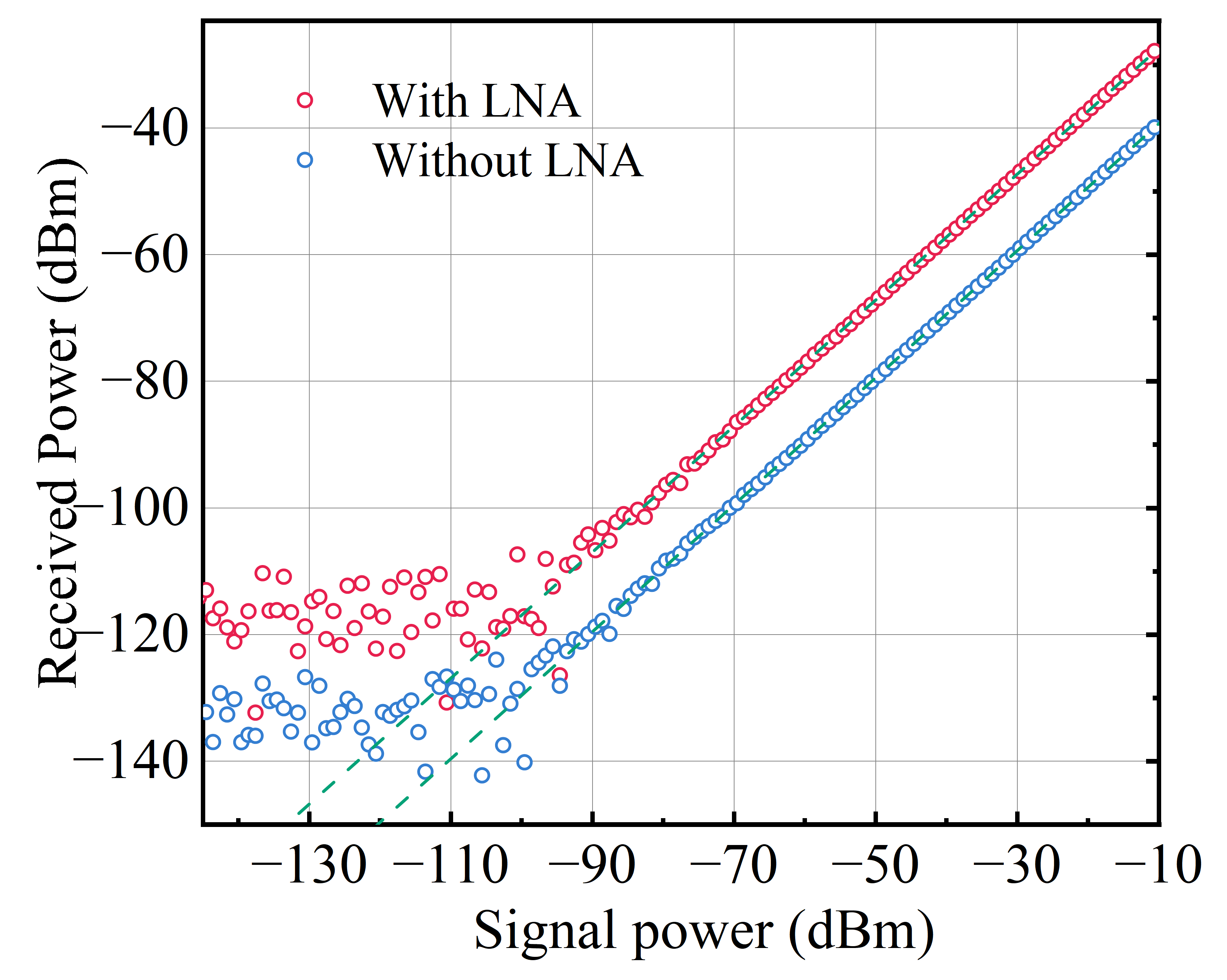}
    \caption{\textbf{Conventional loop-antenna response with and without low-noise amplification.} Received power as a function of signal-source output power measured using a conventional loop antenna with and without a low-noise amplifier (LNA). Red and blue circles denote the results with and without the LNA, respectively. Dashed lines show linear fits to the high-power response. At a received power of -115.76~dBm, the fitted input-power difference is approximately 12.6~dB, indicating the improvement provided by LNA-assisted readout.}
    \label{figS4}
\end{figure}

\FloatBarrier
\subsection*{Supplementary Note 7. Receiving-capability metric calculation}

The receiving-capability metric introduced in the main text jointly
accounts for receiving enhancement, instantaneous receiving bandwidth
and the physical size of the complete receiving feed. It is defined as
\begin{equation}
\Phi
=
\frac{G_{\mathrm{rec}}(BW/f_0)}
{V_{\mathrm{env}}/\lambda_0^3},
\qquad
\Phi_{\mathrm{dB}}
=
10\log_{10}\Phi ,
\end{equation}
where \(G_{\mathrm{rec}}\) is the system-level receiving enhancement,
\(BW\) is the fixed-state 3-dB receiving bandwidth, \(f_0\) is the
centre frequency, \(V_{\mathrm{env}}\) is the effective enclosing
volume of the complete receiving feed, and
\(\lambda_0=c/f_0\).

The receiving enhancement in linear units is related to its
decibel value by
\begin{equation}
G_{\mathrm{rec}}
=
10^{G_{\mathrm{rec,dB}}/10}.
\end{equation}
The corresponding enhancement in decibels was obtained from the
difference between the calibrated minimum detectable input powers:
\begin{equation}
G_{\mathrm{rec,dB}}
=
P_{\min}^{0}
-
P_{\min}^{\mathrm{rec}},
\end{equation}
where \(P_{\min}^{0}\) is the minimum detectable input power of the
bare vapor-cell receiver and \(P_{\min}^{\mathrm{rec}}\) is that of
the receiving configuration being evaluated. Both quantities are
expressed in dBm. For the RAFEF, \(G_{\mathrm{rec}}\) corresponds to
the sensitivity-enhancement factor \(G_{\mathrm c}\) defined in the
main text.

Equivalently, the receiving-capability metric can be evaluated
directly in decibels as
\begin{equation}
\Phi_{\mathrm{dB}}
=
G_{\mathrm{rec,dB}}
+
10\log_{10}\left(\frac{BW}{f_0}\right)
-
10\log_{10}\left(
\frac{V_{\mathrm{env}}}{\lambda_0^3}
\right).
\end{equation}

Because the three receiving structures are approximately planar and
disk-shaped, with thicknesses much smaller than their lateral
dimensions, the radius of the minimum enclosing sphere was
approximated as one-half of the maximum lateral dimension:
\begin{equation}
a\simeq\frac{D}{2}.
\end{equation}
The corresponding effective enclosing volume was therefore estimated
as
\begin{equation}
V_{\mathrm{env}}
\simeq
\frac{4}{3}\pi a^3
=
\frac{\pi D^3}{6}.
\end{equation}
The same enclosing-volume convention was applied to all receiving
feeds. This definition also gives
\begin{equation}
\frac{V_{\mathrm{env}}}{\lambda_0^3}
=
\frac{(ka)^3}{6\pi^2},
\qquad
k=\frac{2\pi}{\lambda_0}.
\end{equation}

All three receiving configurations were evaluated at
\(f_0=93.97~\mathrm{MHz}\). The maximum lateral dimensions were
\(D=130~\mathrm{mm}\) for RAFEF1 and \(D=124~\mathrm{mm}\) for both
RAFEF2 and the loop antenna. The resulting enclosing volumes are
approximately \(1.150\times10^6~\mathrm{mm^3}\) for RAFEF1 and
\(9.983\times10^5~\mathrm{mm^3}\) for RAFEF2 and the loop antenna.
The parameters and resulting receiving-capability values are
summarized in Supplementary Table~\ref{tab:phi_calculation}.

\begin{table}[htbp]
\centering
\caption{\textbf{Parameters used to calculate the receiving-capability metric.}
The receiving enhancement is referenced to the bare vapor-cell
receiver. The volume is estimated using a common enclosing-sphere
convention.}
\label{tab:phi_calculation}
\begin{tabular}{lccccc}
\toprule
Receiver
& \(f_0\) (MHz)
& \(G_{\mathrm{rec,dB}}\) (dB)
& \(BW\) (MHz)
& \(V_{\mathrm{env}}\) (\(\mathrm{mm^3}\))
& \(\Phi_{\mathrm{dB}}\) (dB) \\
\midrule
RAFEF1
& 93.97
& 51.26
& 0.300
& \(1.150\times10^6\)
& 70.81 \\

RAFEF2
& 93.97
& 39.90
& 0.600
& \(9.983\times10^5\)
& 63.07 \\

Loop antenna
& 93.97
& 23.50
& 5.900
& \(9.983\times10^5\)
& 56.60 \\
\bottomrule
\end{tabular}
\end{table}

To provide a size-matched conventional reference for the
frequency-dependent RAFEF measurements, the loop-antenna response was
estimated using geometrical area scaling. For geometrically similar
receiving structures, electromagnetic scaling gives
\(fD\simeq\mathrm{constant}\), where \(D\) is the characteristic
lateral dimension. Assuming that the effective receiving area scales
with the geometrical area, \(A_{\mathrm e}\propto D^2\), gives
\begin{equation}
\frac{A_{\mathrm e}(f)}
{A_{\mathrm e}(f_{\mathrm{ref}})}
\simeq
\left(\frac{f_{\mathrm{ref}}}{f}\right)^2.
\end{equation}
The corresponding area-scaled loop reference is
\begin{equation}
G_{\mathrm{loop,ref}}(f)
=
G_{\mathrm{ref}}
+
20\log_{10}
\left(\frac{f_{\mathrm{ref}}}{f}\right),
\end{equation}
where \(f_{\mathrm{ref}}=78.99~\mathrm{MHz}\) and
\(G_{\mathrm{ref}}=21.66~\mathrm{dB}\) were obtained from the
size-matched loop-antenna measurement. This curve is used only as an
area-scaled conventional reference for comparison with the
frequency-dependent RAFEF measurements.

\FloatBarrier

\FloatBarrier

\bibliography{ref}